\documentstyle[psfig,times]{mn}

\newif\ifAMStwofonts

\def\approxlt{\mathrel{\spose{\lower 3pt\hbox{$\sim$}}
        \raise 2.0pt\hbox{$<$}}}
\def\approxgt{\mathrel{\spose{\lower 3pt\hbox{$\sim$}}
        \raise 2.0pt\hbox{$>$}}}
        
\def\cm{{\rm\thinspace cm}}
\def\ct{{\rm\thinspace ct}}
\def\deg {$^\circ$}
\def\erg{{\rm\thinspace erg}}

\def\km{{\rm\thinspace km}}
\def\kpc{{\rm\thinspace kpc}}

\def\Mpc{{\rm\thinspace Mpc}}

\def\s{{\rm\thinspace s}}

\def\ctps{\hbox{$\ct\s^{-1}\,$}}

\def\ergps{\hbox{$\erg\s^{-1}\,$}}

\def\kmps{\hbox{$\km\s^{-1}\,$}}

\def\pcmsq{\hbox{$\cm^{-2}\,$}}

\def\kmpspMpc{\hbox{$\kmps\Mpc^{-1}$}}

\ifoldfss
  \ifCUPmtlplainloaded \else
    \NewTextAlphabet{textbfit} {cmbxti10} {}
    \NewTextAlphabet{textbfss} {cmssbx10} {}
    \NewMathAlphabet{mathbfit} {cmbxti10} {} 
    \NewMathAlphabet{mathbfss} {cmssbx10} {} 
  \fi
  \ifAMStwofonts
    \ifCUPmtlplainloaded \else
      \NewSymbolFont{upmath} {eurm10}
      \NewSymbolFont{AMSa} {msam10}
      \NewMathSymbol{\upi}     {0}{upmath}{19}
      \NewMathSymbol{\umu}     {0}{upmath}{16}
      \NewMathSymbol{\upartial}{0}{upmath}{40}
      \NewMathSymbol{\leqslant}{3}{AMSa}{36}
      \NewMathSymbol{\geqslant}{3}{AMSa}{3E}

    \fi
  \fi
\fi 

\ifnfssone
  \newmathalphabet{\mathit}
  \addtoversion{normal}{\mathit}{cmr}{m}{it}
  \addtoversion{bold}{\mathit}{cmr}{bx}{it}
  \newmathalphabet{\mathbfit} 
  \addtoversion{normal}{\mathbfit}{cmr}{bx}{it}
  \addtoversion{bold}{\mathbfit}{cmr}{bx}{it}
  \newmathalphabet{\mathbfss} 
  \addtoversion{normal}{\mathbfss}{cmss}{bx}{n}
  \addtoversion{bold}{\mathbfss}{cmss}{bx}{n}
  \ifAMStwofonts
    \ifCUPmtlplainloaded \else
      \UseAMStwoboldmath
      \makeatletter
      \new@mathgroup\upmath@group
      \define@mathgroup\mv@normal\upmath@group{eur}{m}{n}
      \define@mathgroup\mv@bold\upmath@group{eur}{b}{n}
      \edef\UPM{\hexnumber\upmath@group}
      \new@mathgroup\amsa@group
      \define@mathgroup\mv@normal\amsa@group{msa}{m}{n}
      \define@mathgroup\mv@bold\amsa@group{msa}{m}{n}
      \edef\AMSa{\hexnumber\amsa@group}
      \makeatother
      \mathchardef\upi="0\UPM19
      \mathchardef\umu="0\UPM16
      \mathchardef\upartial="0\UPM40
      \mathchardef\leqslant="3\AMSa36
      \mathchardef\geqslant="3\AMSa3E
    \fi
  \fi
\fi 

\ifnfsstwo
  \DeclareMathAlphabet{\mathbfit}{OT1}{cmr}{bx}{it}
  \SetMathAlphabet\mathbfit{bold}{OT1}{cmr}{bx}{it}
  \DeclareMathAlphabet{\mathbfss}{OT1}{cmss}{bx}{n}
  \SetMathAlphabet\mathbfss{bold}{OT1}{cmss}{bx}{n}
  \ifAMStwofonts
    \ifCUPmtlplainloaded \else
      \DeclareSymbolFont{UPM}{U}{eur}{m}{n}
      \SetSymbolFont{UPM}{bold}{U}{eur}{b}{n}
      \DeclareSymbolFont{AMSa}{U}{msa}{m}{n}
      \DeclareMathSymbol{\upi}{0}{UPM}{"19}
      \DeclareMathSymbol{\umu}{0}{UPM}{"16}
      \DeclareMathSymbol{\upartial}{0}{UPM}{"40}
      \DeclareMathSymbol{\leqslant}{3}{AMSa}{"36}
      \DeclareMathSymbol{\geqslant}{3}{AMSa}{"3E}
    \fi
  \fi
\fi 

\ifCUPmtlplainloaded \else
  \ifAMStwofonts \else 
    \def\upi{\pi}
    \def\umu{\mu}
    \def\upartial{\partial}
  \fi
\fi

\title{ Extended X-ray emission around four 3C quasars at $0.55<z<0.75$ observed with {\sl Chandra} }
\author[C. S. Crawford \& A.C.Fabian ]
       {C. S. Crawford  and A. C. Fabian  \\
        Institute of Astronomy, Madingley Road, Cambridge CB3 0HA}

\date{Submitted 2 August 2002: revised version submitted 11 Oct 2002 }

\pubyear{2002}

\begin{document}

\maketitle

\label{firstpage}

\begin{abstract}
We report on the detection of a soft spatially-extended component of
X-ray emission around four intermediate-redshift 3C quasars observed
with {\sl Chandra}: 3C254, 3C263, 3C275.1 and 3C281. The bolometric
luminosity of this emission ranges over 0.3--1.6$\times10^{44}$\ergps,
and extends to lengthscales of over 350\kpc\ at the redshift of the
quasar. The X-rays are most likely thermal emission from the
intracluster medium of a cluster of galaxies around each quasar, which
provides the working surface for the powerful radio lobes. Some X-ray
emission is also seen to be associated with the radio plasma. 
\end{abstract}

\begin{keywords}
galaxies: clusters: general -- quasars: general -- X-rays: galaxies
\end{keywords}

\section{Introduction}

There has been accumulating evidence for many years that
intermediate-redshift, radio-loud quasars reside in strongly clustered
environments. The optical field galaxy counts around some radio-loud
quasars (Wold et al 2000; Ellingson, Yee \& Green 1991; Yee \& Green
1987), and the high gas pressures inferred in the extended
emission-line nebulae (Crawford \& Vanderriest 1997, 2000; Durret et
al 1994; Bremer et al 1992) suggest that they lie in clusters
typically of Abell class 0 or richer. Attempts to detect
spatially-extended and thermal X-ray emission from any cluster
component were made using the {\sl ROSAT} satellite (Crawford et al
1999; Hardcastle \& Worrall 1999). An extended component was found
around five intermediate-redshift 3C quasars (3C215, 3C254, 3C275.1,
3C281 and 3C334), having bolometric luminosities ranging over
2-17$\times10^{44}$\ergps, and characteristic lengthscales of
25-265\kpc. The detection of the extended X-ray emission from {\sl
ROSAT HRI} data, however, was complicated by the need to remove the
effects of spacecraft wobble before analysis.  In this paper we
revisit the environment of three of these 3C quasars (3C254, 3C275.1
and 3C281) and that of a further intermediate-redshift quasar (3C263)
using our own pointed, and archival, {\sl Chandra} observations.

We assume a cosmology of H$_0=50$\kmpspMpc and q$_0=0.5$ throughout this
paper.

\section{Observations}
3C281 was observed by {\sl Chandra} with the ACIS-S on 2001 May 30
using the FAINT telemetry mode, with the source position
offset from the aimpoint of chip 7 by just over an arcminute. The
observation was free from background flaring events, so we
extracted data from the total exposure of 15.85~ksec.

We retrieved ACIS-S data from the {\sl Chandra} archive on three other
quasars at intermediate redshifts: 3C254, 3C263 and 3C275.1. Despite being a
spectroscopic observation, we also examined the zeroth order image of
the lower-redshift QSO 1821+643 which is known to lie in a cluster.
The environment of this quasar has been previously studied by
Fang et al (2002), and we include it to enable comparison of our
method to a result in the literature. A log of observations is given in
Table~\ref{tab:obs}.
There was no obvious flaring activity in the soft band on chip 7 for
the observations of 3C254, 3C275.1 and 1821+643, so we took the data
from the full exposure in each case. The observation of 3C263 showed a
high background countrate at the start of the observation, so we used
data from only the latter two-thirds of the exposure.

\begin{figure}
\hbox{
\psfig{figure=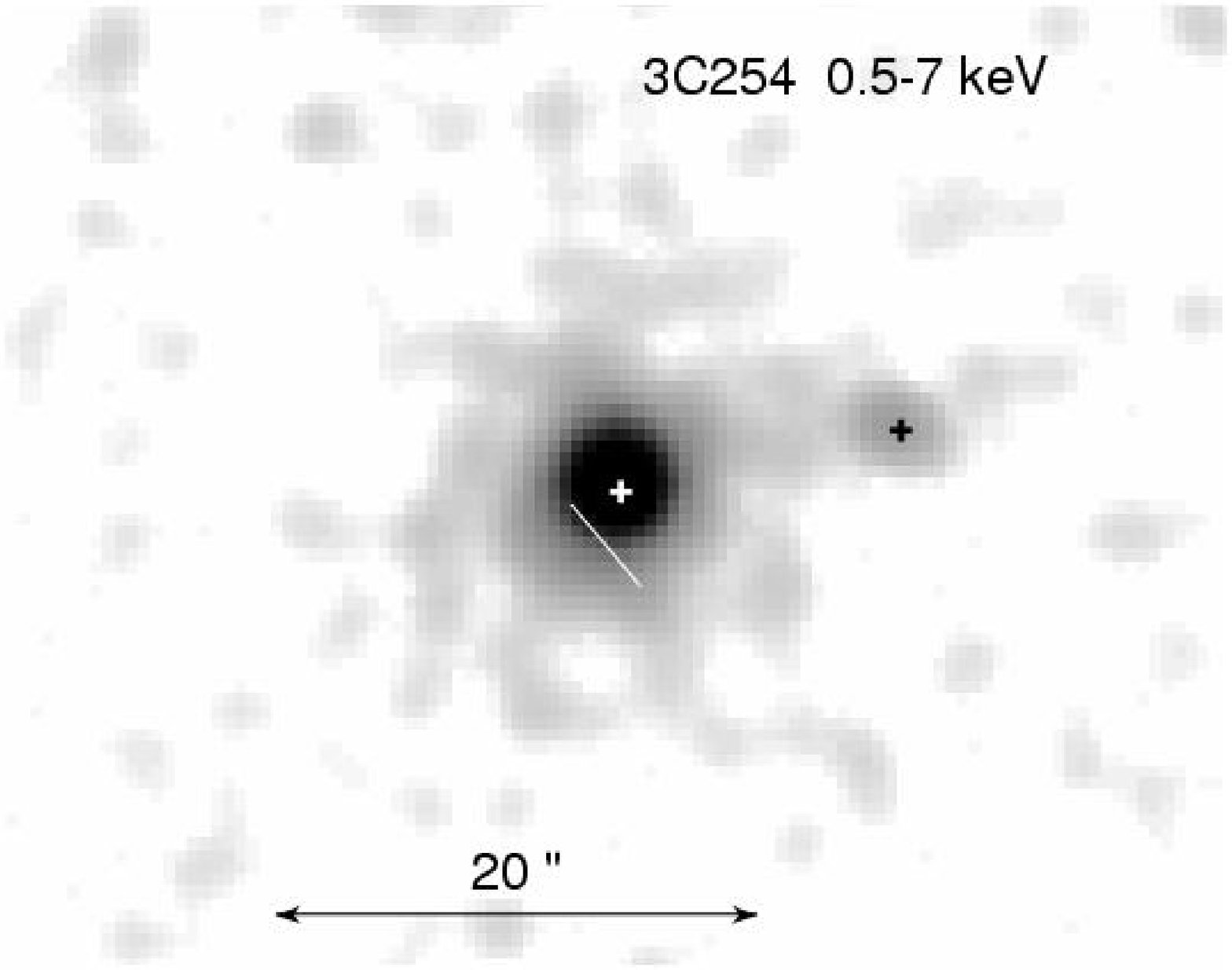,width=0.33\textwidth,angle=0}
\psfig{figure=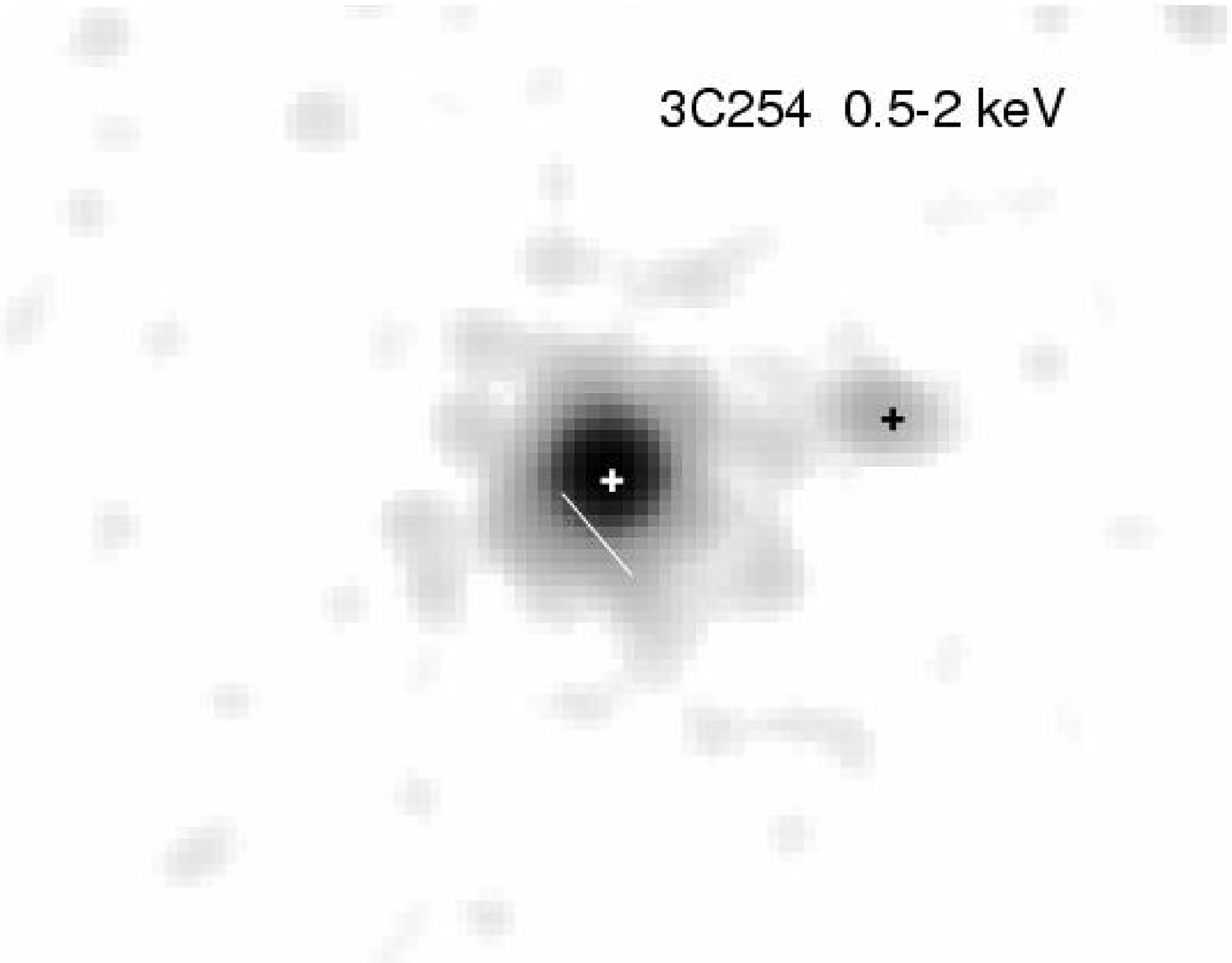,width=0.33\textwidth,angle=0}
\psfig{figure=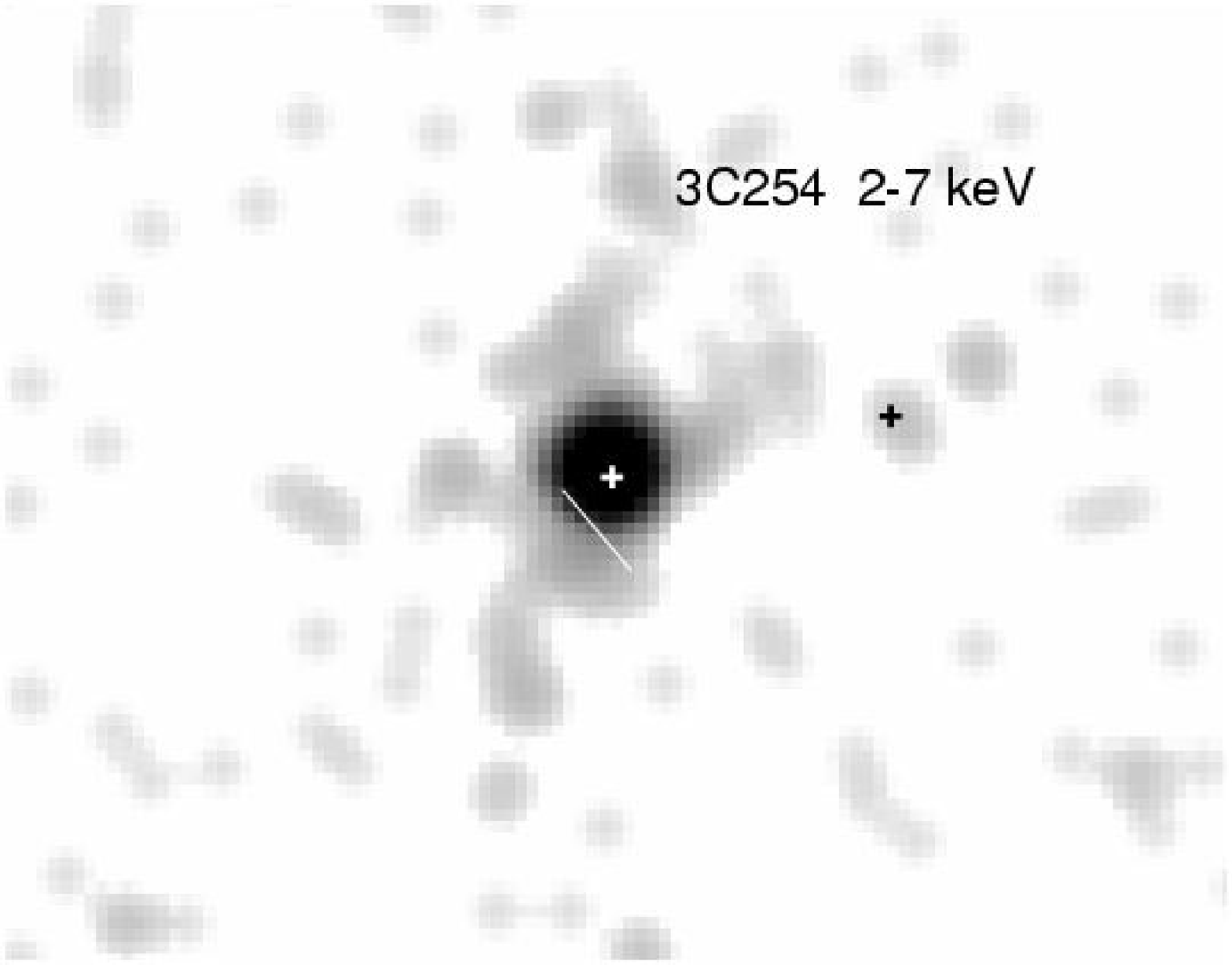,width=0.33\textwidth,angle=0} 
}
\hbox{
\psfig{figure=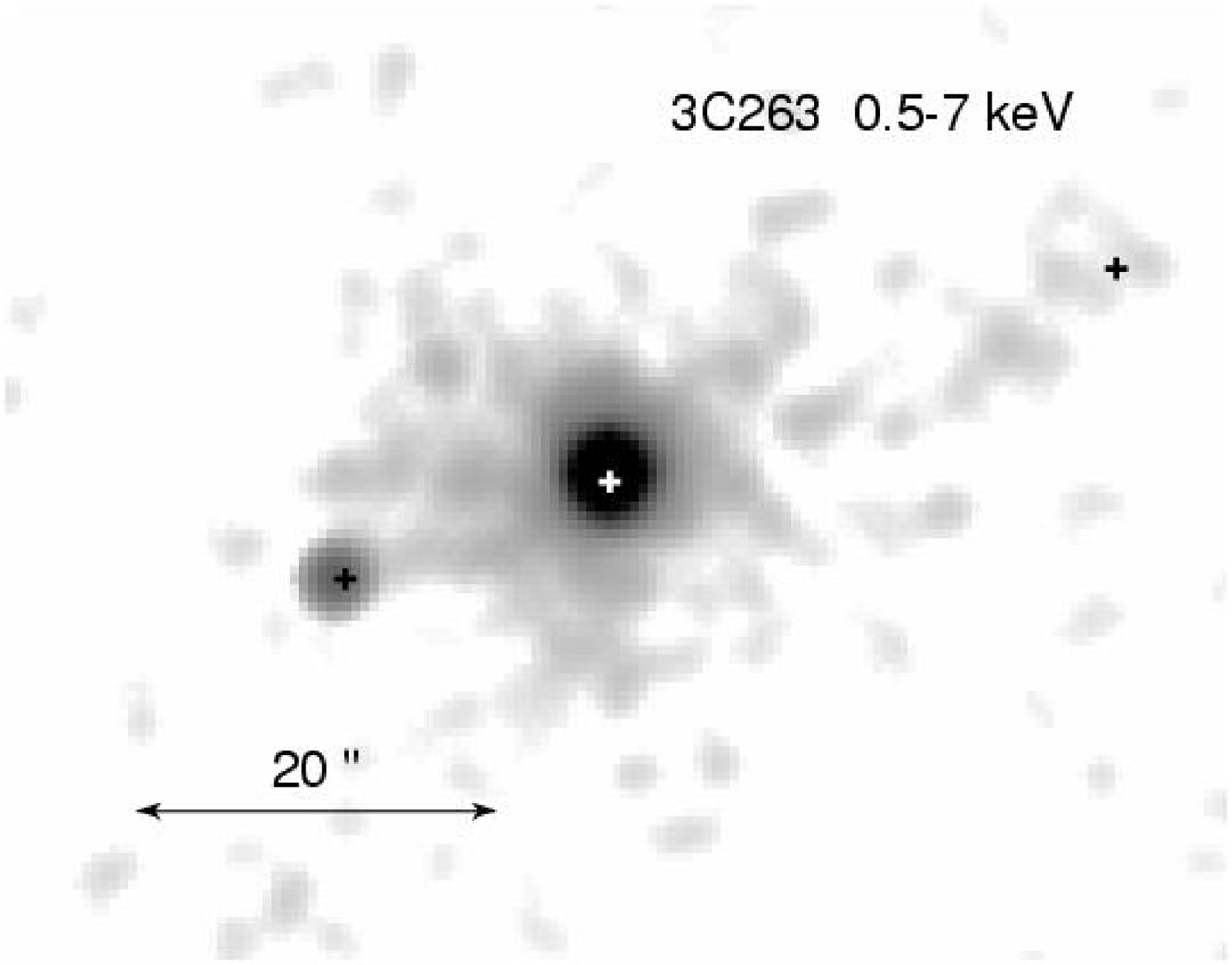,width=0.33\textwidth,angle=0}
\psfig{figure=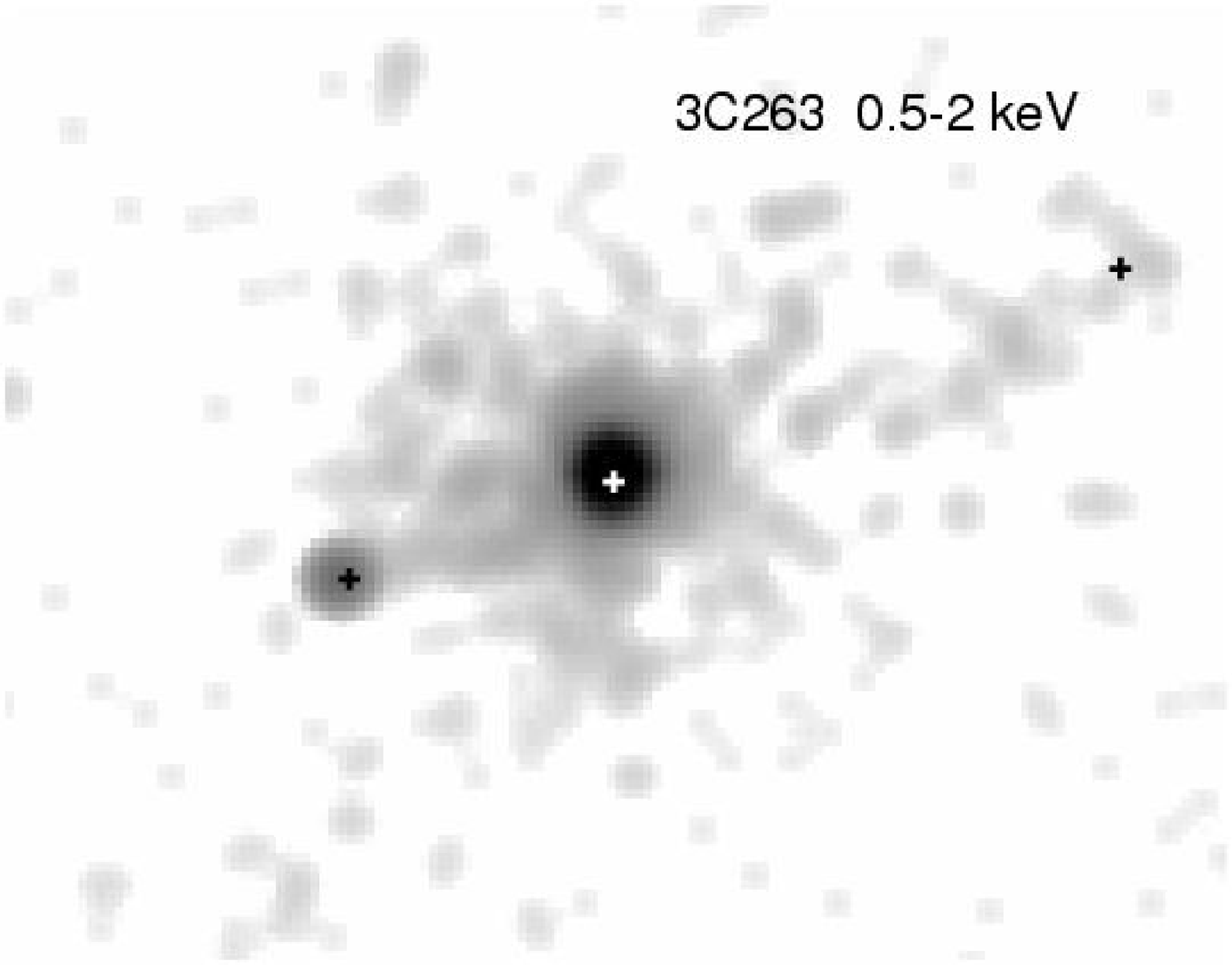,width=0.33\textwidth,angle=0}
\psfig{figure=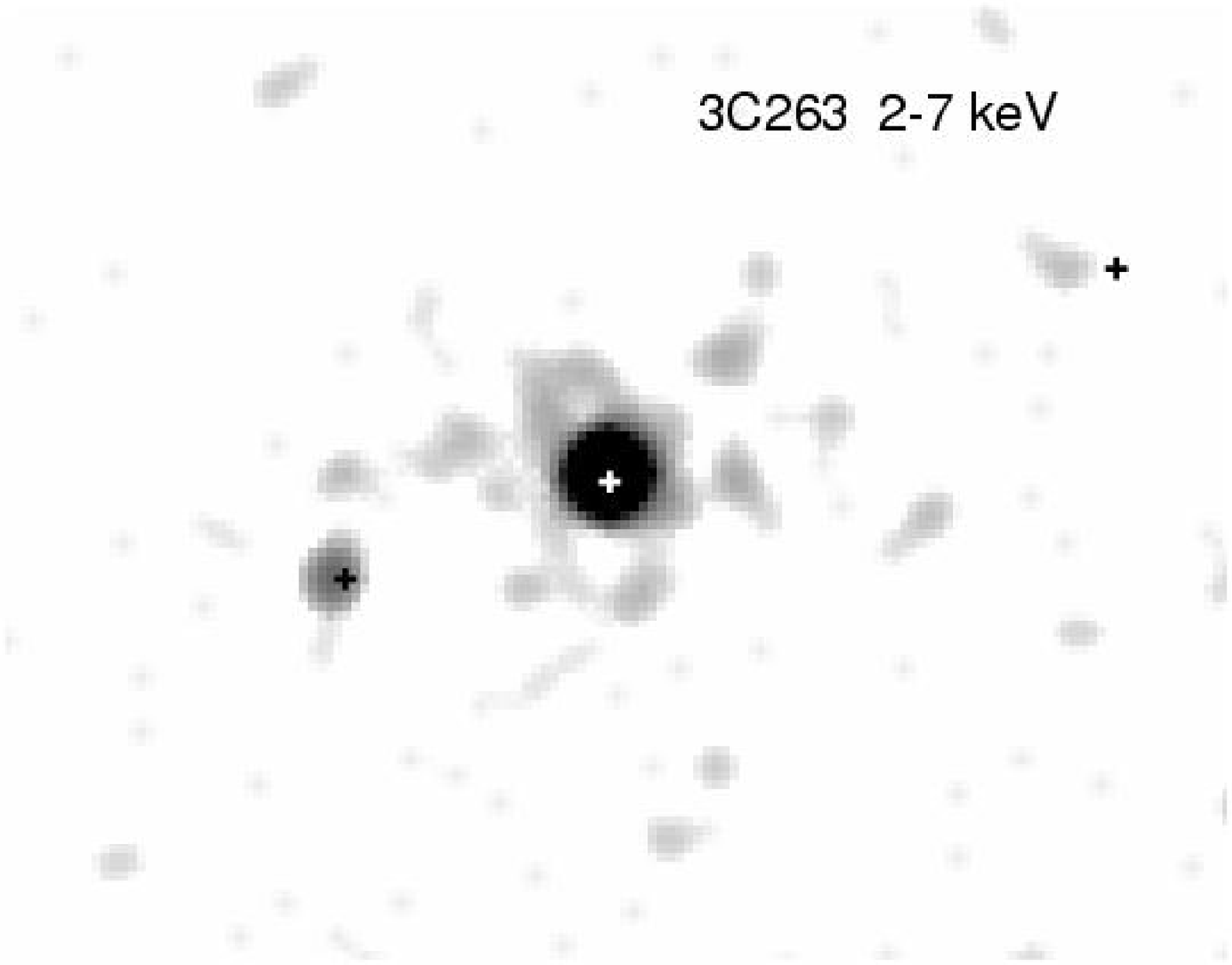,width=0.33\textwidth,angle=0} }
\hbox{
\psfig{figure=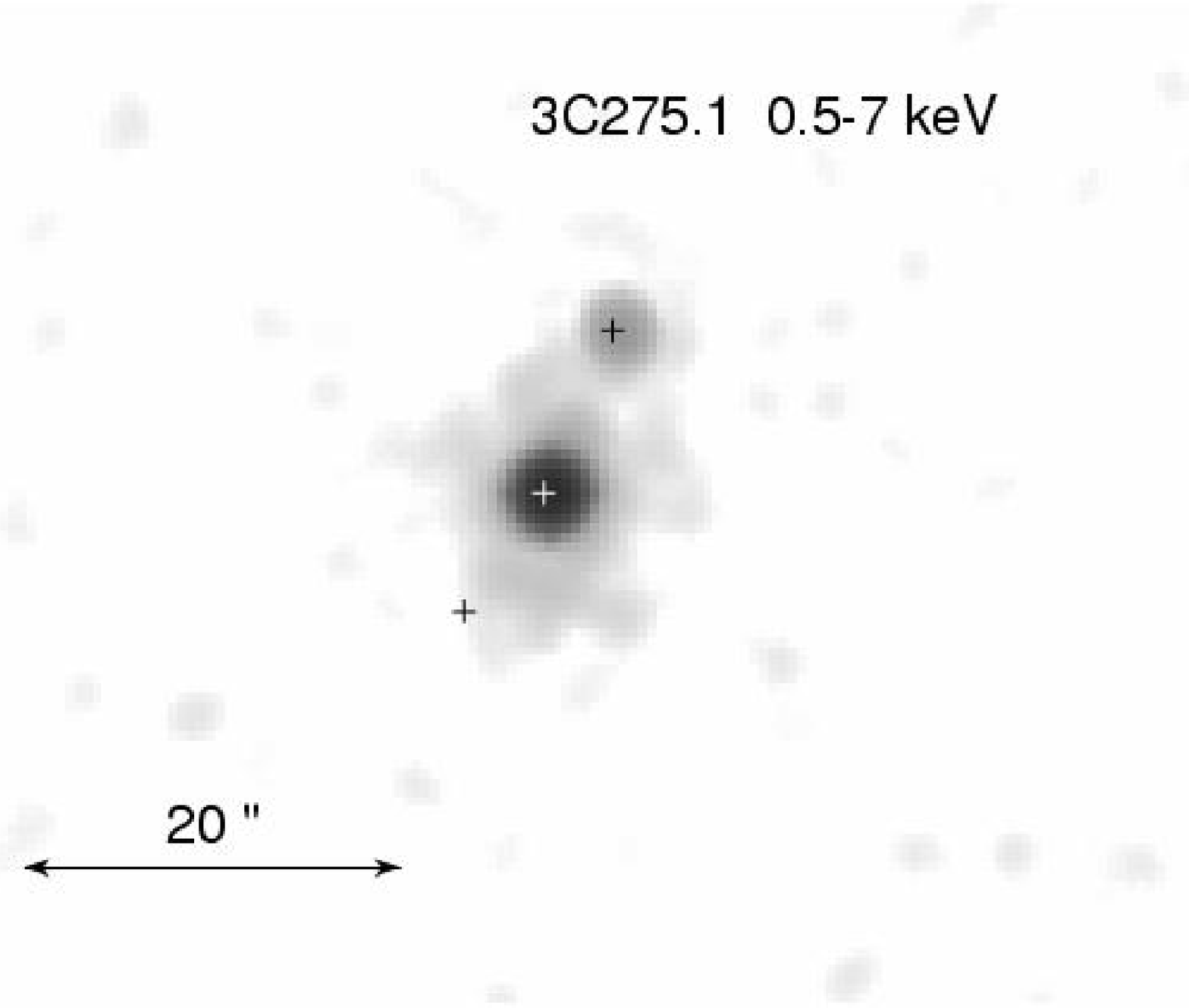,width=0.33\textwidth,angle=0}
\psfig{figure=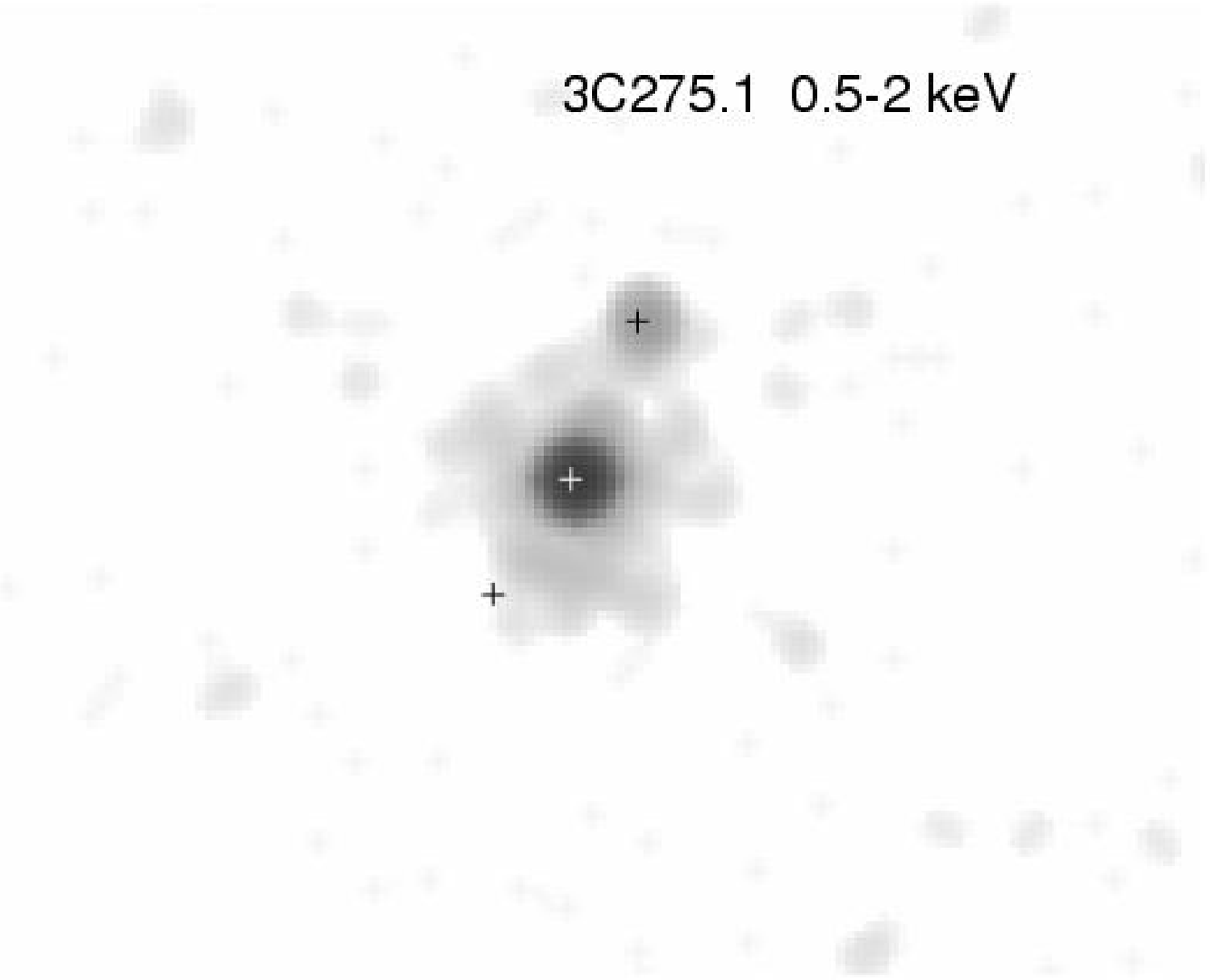,width=0.33\textwidth,angle=0}
\psfig{figure=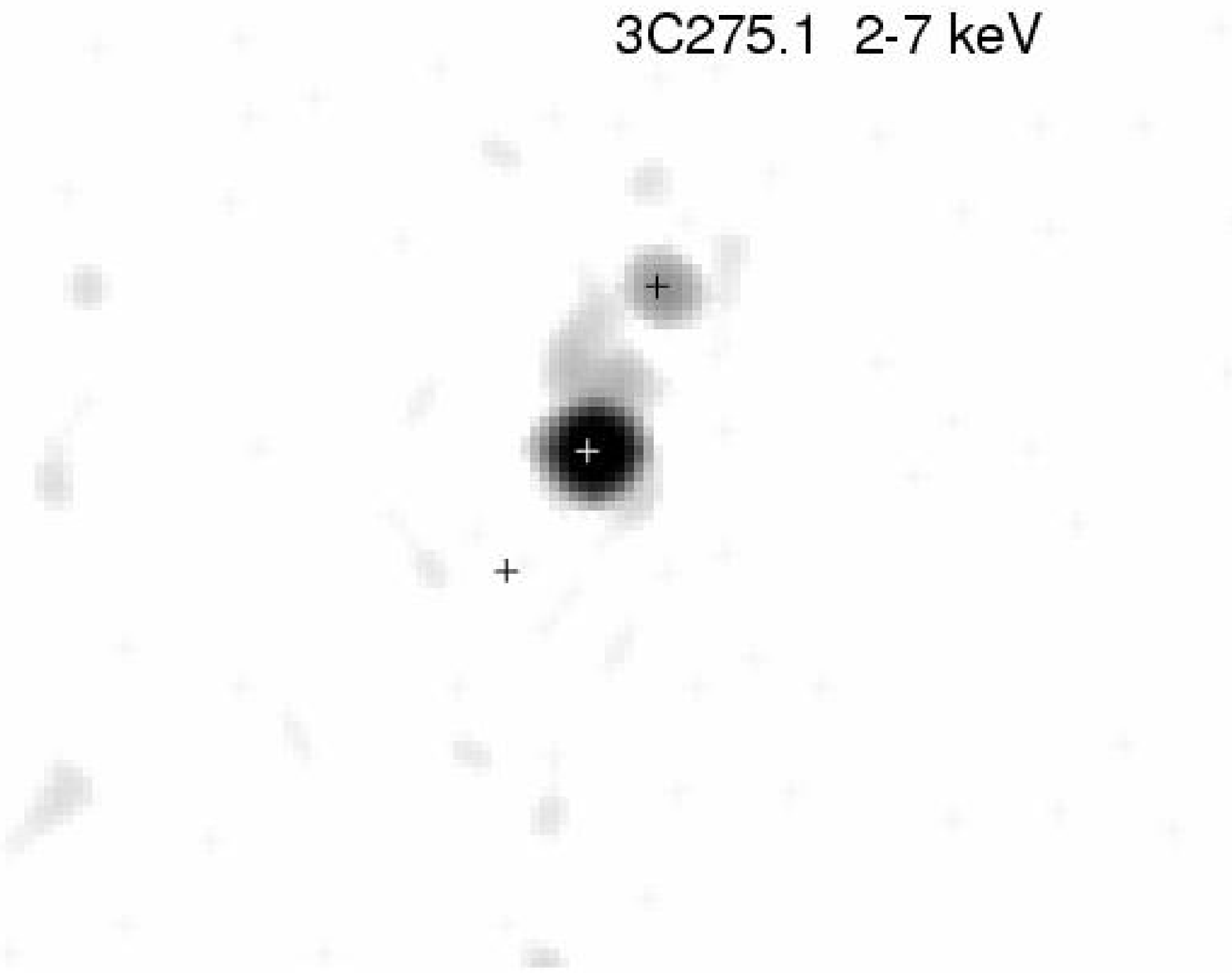,width=0.33\textwidth,angle=0} }
\hbox{
\psfig{figure=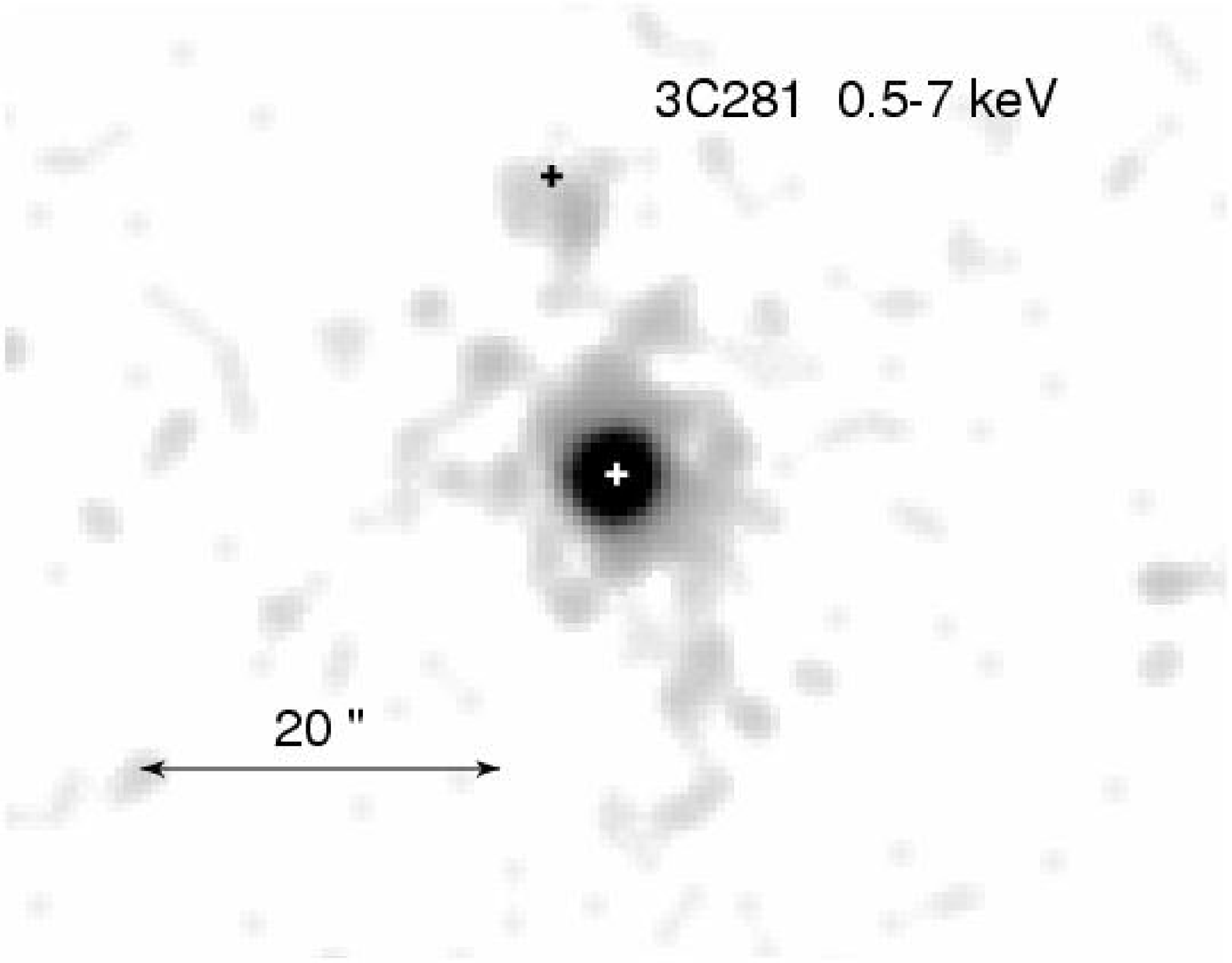,width=0.33\textwidth,angle=0}
\psfig{figure=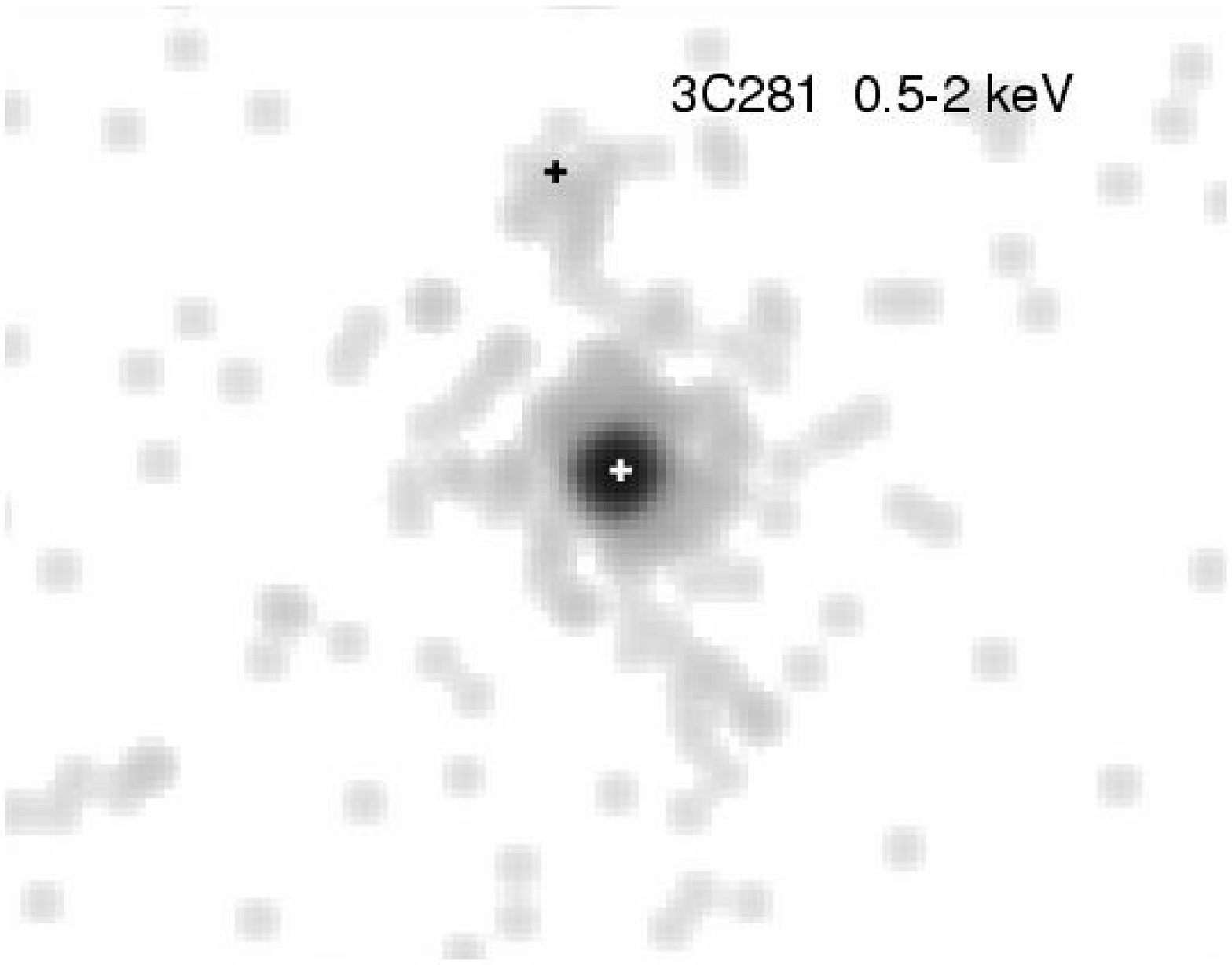,width=0.33\textwidth,angle=0}
\psfig{figure=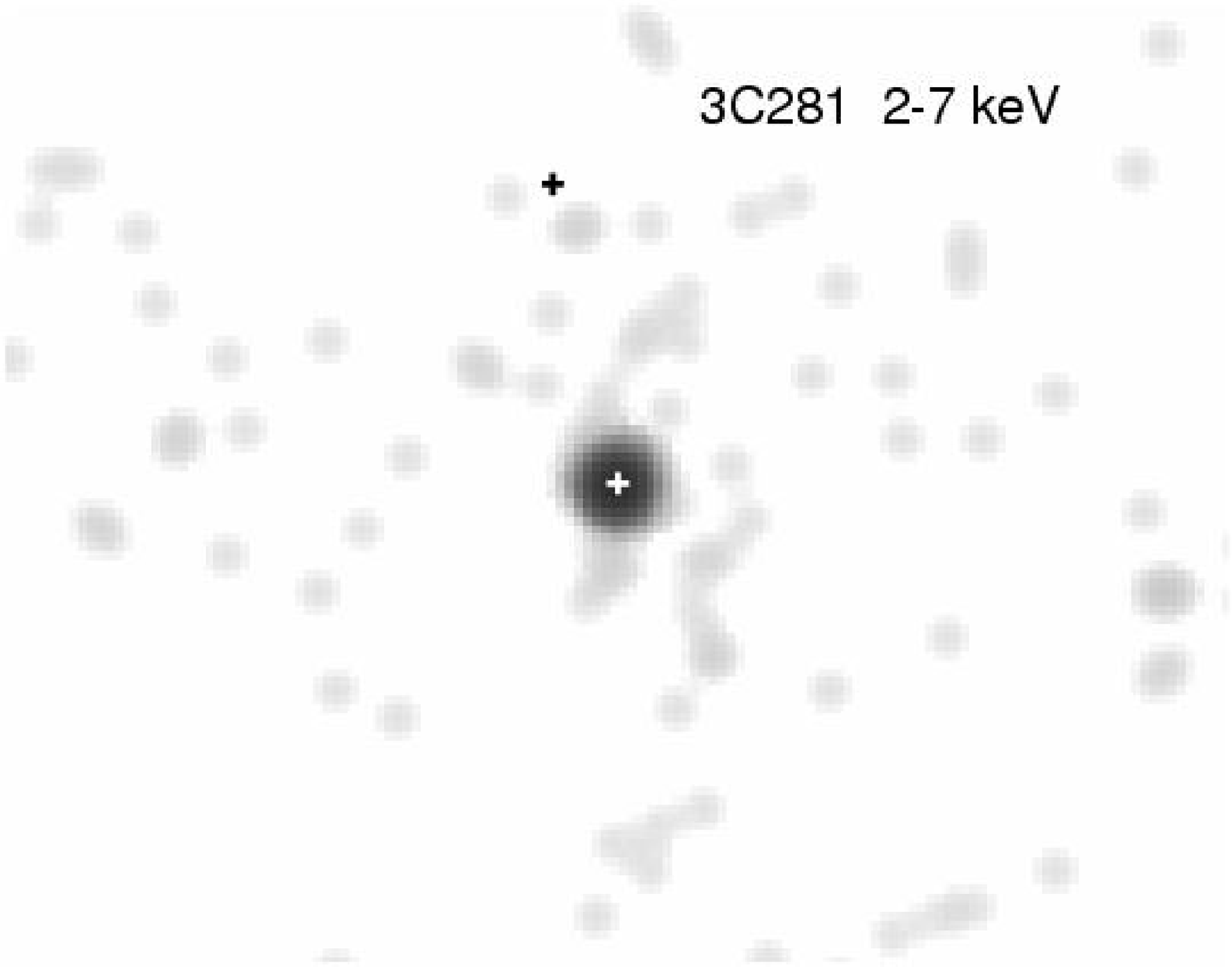,width=0.33\textwidth,angle=0} }

\caption{ \label{fig:ximgs}
Images of the {\sl Chandra} X-ray emission from the
intermediate-redshift 3C quasars in the 0.5-7~keV (left), 0.5-2~keV
(middle) and 2-7~keV bands (right). The data have been smoothed with a
Gaussian of 2 pixels (ie 1 arcsec). The positions of the principle
radio source components are shown on the 0.5-7~keV maps, as measured
from Owen \& Puschell 1984 (for 3C254); Hutchings et al 1998 (for 3C263,
3C275.1 and 3C281). The radio nucleus is marked as a white cross in
each case, with the hotspots of radio lobes marked as black crosses.
The white line in the picture of 3C254 shows the position and
alignment of the highly distorted radio lobe to the south-east of the
quasar nucleus. }
\end {figure}

X-ray emission from the quasar is easily detected from all five
sources; images of the four 3C quasars in each of three energy bands
0.5-7~keV, 0.5-2~keV (soft) and 2-7~keV (hard) are shown in Figure~\ref{fig:ximgs}. On
an initial visual inspection, all the quasars appear to have a diffuse
halo of spatially extended emission surrounding the central quasar
nucleus in the soft (0.5-2~keV) band. Each also has an offset X-ray
source at the position of the more prominent radio hotspot: to the
west of 3C254, the south-east of 3C263, the north-west of 3C275.1 and
to the north of 3C281. The images of 3C254 also suggest extended
emission slightly beyond, but still following the line of the highly
distorted radio lobe to the south-east of the nucleus. We extract the
net counts from a circular aperture over each of the hotspots, using
an annular region at large radii around the quasar (see later) for
estimation of the background. The counts from, and the luminosities
inferred for these \lq X-ray hotspots' (assuming a $\Gamma=2$
power-law spectrum at the same redshift as the quasar, and with only
Galactic absorption) are presented in Table~\ref{tab:hots}.

We are particularly interested in detecting any possible extended
component of the X-ray emission around these quasars. It is thus
important to assess whether the nuclear quasar emission suffers from a
large degree of pileup, as this will distort the point-spread function
(PSF) of the nucleus: if pileup is significant in the core, the centre
of the nuclear PSF will be artificially suppressed relative to its
wings. We used DMEXTRACT to estimate the number of counts from the
X-ray source within a radius of 10 pixels (5 arcsec) centred on each
quasar nucleus, without background-subtraction. Most of the quasar
observations have a frame-time of 3.2s, except for the data from
3C275.1, which is sub-arrayed (to one-half the chip size, and so has a
frame-time of 1.8s). We list the counts/frame-time from these central
regions in Table~\ref{tab:obs}; the higher values of 0.6-0.65
correspond to a modest pile-up fraction of around 25 per cent at the
core of the X-ray source (CXC Handbook). 

We extracted surface brightness profiles of the X-ray emission
associated with each of the quasars using the CIAO software package
DMEXTRACT. As we are concerned with testing for the presence of
extended, thermal emission, we produce the azimuthal profiles only in
the soft (0.5-2~keV) energy band. The counts were extracted from
annuli centred on the quasar nuclear emission, extending out to a
radius of 50 arcsec for the four 3C quasars. The annuli were of width
0.5 arcsec out to a 10 arcsec radius, 1 arcsec between 10 and 15
arcsec, 1.5 arcsec between 15 and 30 arcsec, and 2.5 arcsec beyond.
The smoothed images also suggest the presence of jetted soft X-ray
emission along the radio axis between the radio lobes in at least
3C263 and 3C281. To be conservative, we excluded X-ray emission from
sectors spatially corresponding with the radio source in all four 3C
quasars.  Specifically: the sector between 270 and 300\deg~ (where the
angles increase anticlockwise from North) out to a
radius of 15 arcsec for the western arm of the radio source in 3C254,
and from 90 to 110\deg~ out to 5 arcsec for the distorted SE lobe;
between 104 and 121\deg~ out to 19.5 arcsec and between 288 and 298
\deg~ out  to 32.5 arcsec for the radio source arms in 3C263;
between 168--192\deg~ and 325--345\deg~ out to 11 arcsec radius for
3C275.1; and between 4--24\deg~ out to a 20 arcsec radius, and between
185-205\deg~ out to 33 arcsec for 3C281. A background source directly
to the E of 3C254 at 30-34 arcsec was also excluded.  The background
counts in each case were estimated from a larger concentric annular
region centred on the quasar: between radii of 65--88 arcsec for
3C254, 60--105 arcsec for 3C263, 67--85 arcsec for 3C275.1 and 60--90
arcsec for 3C281. The profile for 1821+643 was extracted out to a
radius of 75 arcsec, and the background estimated from a large,
slightly offset region. Within these radii, the counts in the readout
streaks were negligible for all the 3C quasars, and were excluded from
the profile of 1821+643.

%

\section{Determination of a {\sl Chandra} point spread function}

\subsection{Model PSFs}

We used CIAO/MKPSF with the PSF library available from the CXC to
estimate a two-dimensional model of the PSF for each of our target
quasars. The PSF was created at the position of the quasar on the
detector, in seven discreet bands that sampled the energy range
between 0.5 and 2~keV at 0.25~keV intervals. This sampling is required
as the higher energy PSFs are broader than that at lower energy. These
separate energy PSFs were coadded, with relative normalizations of the
profile in each 0.25~keV interval scaled in proportion to the shape of
the incident spectrum of the quasar (ie onto the mirror). This thus
provides a composite theoretical estimate of the PSF of the quasar
nucleus at the observed chip position.

The incident quasar continuum shape (used for scaling of the energy
bands) was estimated using a background-subtracted nuclear spectrum
taken from within the central 3.5 arcsec radius of the quasar X-ray
centroid (except for 1821+643, where we instead used the incident
spectrum published by Fang et al 2002). The nuclear spectra were
fit by a simple absorbed power-law model, with the absorption fixed to
be the Galactic column density in the direction of the quasar. The
nuclear spectra of the four 3C quasars were fit (sometimes not so
well, except for 3C275.1) by flat power law models
($\Gamma\sim1.2-1.5$). The slightly flatter slope of these models
compared to the canonical power-law fits observed for quasars (eg Sambruna, Eracleous
\& Mushotzky 1999) supports the
inference that the observations are mildly piled up. However, the
approximation of the incident continuum is sufficiently good for the
normalization of the model PSFs; we found from experimentation that
the major variance in the PSF output from MKPSF was the position of
the source on the chip, rather than from the spectral slopes assumed
for the incident continuum (see later). 

\begin{figure}
\psfig{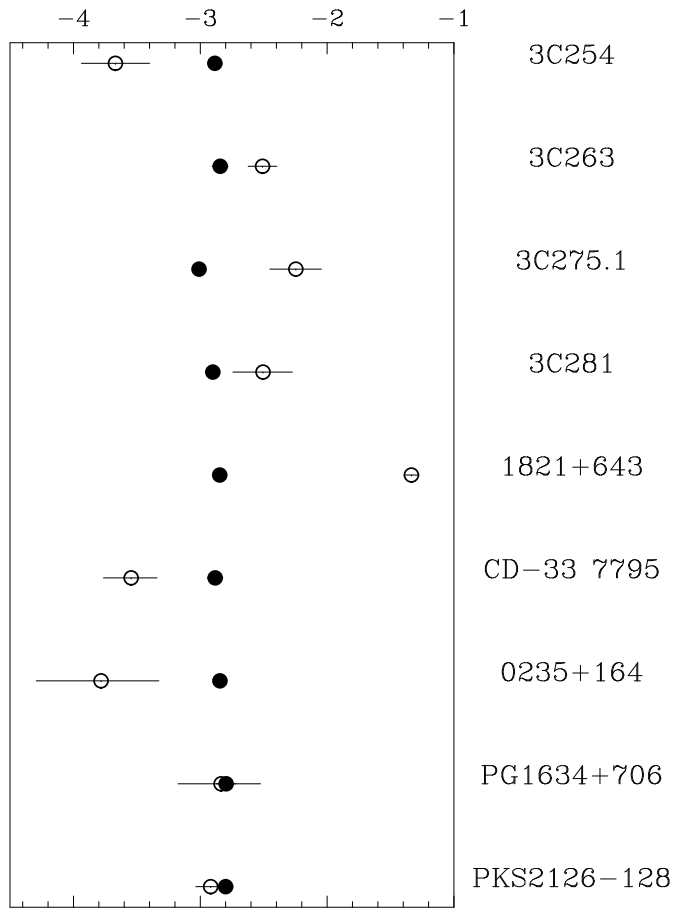}
\caption {\label{fig:fits} The measured slope of the point spread function over radii
2-10 arcsec. The slopes of the MKPSF model appropriate for each quasar
observation is shown by a solid marker and that of the PSF of the
actual data is shown by an open circle marker. The data for four of
the target quasars (ie except 3C254) show shallower slopes for the
data than the model PSFs, whereas the converse is true for the four
\lq control' observations.}
\end{figure}

The MKPSF creates model PSFs out to radii of 10 arcsec, and those
produced for the five quasars vary slightly.  The possibility of
modest pile-up in the central pixels leads us to concentrate on
comparing the more extended components of the PSF. Beyond 2 arcsec,
the slope of the PSF is almost identical for 3C263 and 1821+643
(Figure~\ref{fig:fits}); similarly the slopes for 3C254 and 3C281 are
also very close to each other, although steeper than for the other two
quasars. The model for 3C275.1 has by far the steepest profile. We
ascribe this variation in the extended PSF slope as predominantly due
to the different positions of the sources on the chip: the positions
of 3C254 and 3C281 are very close to one another, whereas those of
3C263 and 1821+643 are offset. We confirm this interpretation by
re-creating the two model PSFs for 3C263 and 1821+643, with the
correct incident spectra but now supposing the quasars were located at
the same position on the chip as 3C254 and 3C281. The resulting model
profiles almost exactly match those of 3C254 and 3C281. However, as
MKPSF itself warns, these model PSFs may be insufficient for detailed
analysis and deconvolution of source profiles. We thus test
their worth against empirical determinations of the PSF from archival
{\sl Chandra} data.


\subsection{Empirical PSFs}

We searched the archive (as of 2002 May) for publicly-available ACIS-S
observations of both normal stars and distant \lq normal' active
galactic nuclei, excluding those without sufficient counts for a
reliable PSF. We also excluded observations where the source was far
from the positions of any of the target quasars on the chip. There was
one stellar observation fulfilling these criteria, and three AGN
observations. The most important datasets for this purpose, however,
are the two with the most signal in the source: the long observation
of the $z=1.34$ quasar PG1634+706 created by merging six different
archival datasets (obsID's of 47, 62, 69-71, 1269) using the CIAO
script MERGE\_ALL, and the observation of the $z=3.27$ quasar
PKS~2126-128. These quasars are also at sufficiently high redshift
that even if they lie in a clustered environment themselves, we do not
expect to detect any extended component of X-ray emission. The details
of the archival datasets used are listed in Table~\ref{tab:emp}. The
counts/frame-time we estimate from each of these observations do not
exceed those of our quasar targets, so any pile-up fraction from these
observations should be comparable or less than in our target quasars.

We used DMEXTRACT to extract a PSF in the 0.5-2~keV band centred on
each of these X-ray sources, with background taken either from a large
annulus further out around each source, or from a neighbouring region
free of background sources. The PSF annuli for PG1634+706 and
PKS~2126-128 had a narrow angle omitted to either side of the quasar
so as to exclude the readout stripes. For comparison, we also created
a model PSF for each of the four \lq control' observations, for the
same incident spectrum and position on the chip as the source (as
described in the previous section). We again measured the slopes of
the profile of the data and its appropriate PSF model at large radii
(2-10 arcsec; where the profile is flatter) by a power-law fit.
Figure~\ref{fig:fits} shows that in all these empirical cases the
model PSF created for an object has a shallower slope than that taken
from the actual data.

\begin{figure}
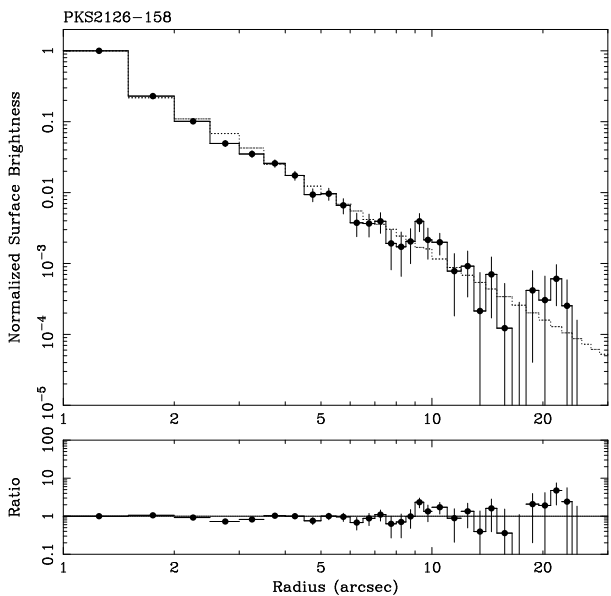

\vbox{
\psfig{figure=new376.ps,width=0.45\textwidth,angle=270}
\psfig{figure=rat376.ps,width=0.45\textwidth,angle=270}}
\caption{ \label{fig:empsfs} The surface brightness profile of the
data (solid line and solid circle markers) and extrapolated PSF model
(dotted line) for the \lq control' quasar PKS~2126-128 (upper panel).
Both profiles have been normalized at the bin at a radius of 1.25
arcsec. The lower panel shows the ratio of the quasar profile to
model, demonstrating that the quasar shows a net deficit of data
counts compared to the model PSF.  }
\end{figure}

We thus will follow a conservative approach in approximating the {\sl
Chandra} PSF of the nuclear X-ray light from each quasar by its MKPSF
model PSF. However, we need to extend this PSF model out beyond 10
arcsec, and after experimentation find a simple and sufficient method
is to extrapolate each PSF following the power-law slope fit to the
model between 5-10 arcsec. This extrapolation is good for the good for
the \lq control' observations (eg PKS2126-158 in Fig~\ref{fig:empsfs};
note that its profile can be extracted only out to 30 arcsec, as this
observation has been made on a subarray of the chip).

\section{Results}
\begin{figure}
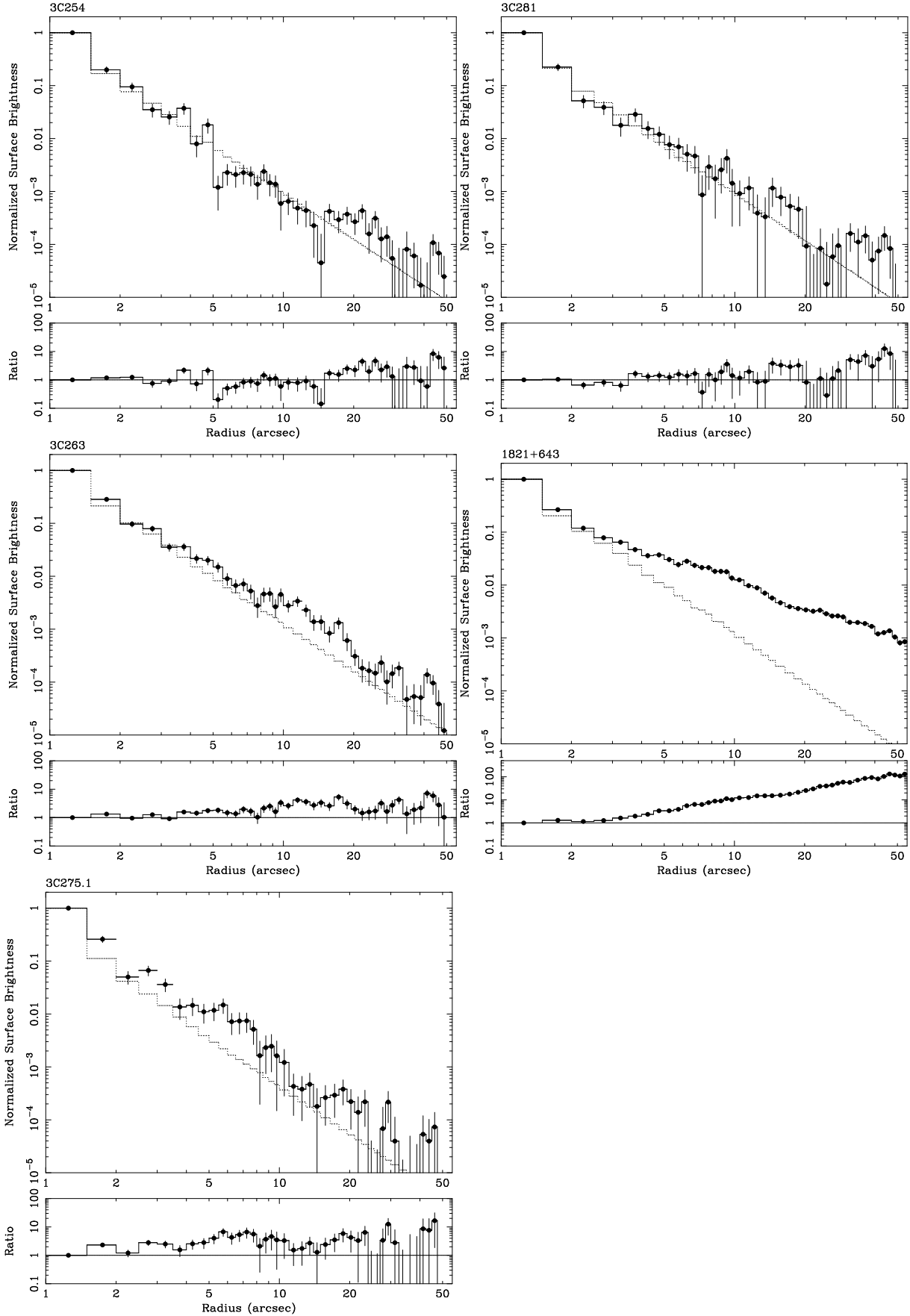

\vbox{
\hbox{
\vbox{
\psfig{figure=new254vsmods.ps,width=0.45\textwidth,angle=270}
\psfig{figure=rat254.ps,width=0.45\textwidth,angle=270}}
\vbox{
\psfig{figure=new281vsmods.ps,width=0.45\textwidth,angle=270}
\psfig{figure=rat281.ps,width=0.45\textwidth,angle=270}}
}
\hbox{
\vbox{
\psfig{figure=new263vsmods.ps,width=0.45\textwidth,angle=270}
\psfig{figure=rat263.ps,width=0.45\textwidth,angle=270}}
\vbox{
\psfig{figure=new1821vsmods.ps,width=0.45\textwidth,angle=270}
\psfig{figure=rat1821.ps,width=0.45\textwidth,angle=270}}
}
\vbox{
\psfig{figure=new275vsmods.ps,width=0.45\textwidth,angle=270}
\psfig{figure=rat275.ps,width=0.45\textwidth,angle=270}}
}
\caption{ \label{fig:psfs} The 
surface brightness profile of each of the five target quasars. 
The data are drawn by solid circle markers and a solid line, and each
quasar profile has been compared to its own model (shown as a dotted
line), normalized at a radius of
1.25 arcsec. Underneath each plot is the ratio between the surface
brightness profile and the model PSF, with a horizontal line drawn at
the value of unity.  }
\end{figure}

Figure~\ref{fig:fits} demonstrates that the model PSFs are shallower
than the surface brightness profiles of the four \lq control'
observations over the radial range of 2-10 arcsec. The converse is
true for all but one (3C254) of the intermediate redshift quasars,
which show an obviously shallower slope than the model PSFs. In order
to quantify this difference in PSFs better, and to obtain an estimate
of the \lq excess' luminosity, we compare the model and
actual surface brightness distributions in detail for each quasar. To
avoid complications from pile-up, we normalize the profiles at the
1.25$\pm$0.25~arcsec radial bin, by which point the quasar flux has
dropped by a factor of 20-30. In Figures~\ref{fig:empsfs} and
\ref{fig:psfs} we show the surface brightness profiles of each of
the five target quasars and the brightest control quasar PKS~2126-128,
along with the appropriate model PSF for each. The ratio of the
profile of the data to that of the normalized model are also shown for
each case in the lower panel.

We add up the excess counts (ie the excess of the data over the model)
 within two annuli: 1.5--10 arcsec, the region for which the CIAO PSF has
 been created, outside of any possible piled up contribution from the
 nucleus; and 10--50 arcsec (10--75 arcsec for 1821+643), using the
 extrapolated model PSF. [To obtain the total excess counts it is
 necessary to scale back up by the annular area used in each ring, as
 well as the factor used to normalize the profiles at 1.25 arcsec.]
 The excess counts and their 1-$\sigma$ uncertainties are listed in
 Table~\ref{tab:excess}. PKS~2126-128 shows a net deficit of counts on
 both scales, which supports our approach of using model PSFs scaled
 at a radius of 1.25~arcsec as indeed a conservative approach. We
 also estimate a significance for the detection of excess counts by finding
 the 1-$\sigma$ uncertainty predicted from the model PSF counts in the
same area; these uncertainties are listed
 in square brackets next to each value of excess counts.

As can be seen in Figure~\ref{fig:psfs} and Table~\ref{tab:excess},
three of the target quasars (3C263, 3C275.1 and 1821+643) show a clear
and very significant excess of counts over the scaled PSF at both the 
1.5--10 and 10--50 (or 75) arcsec scale. 3C254 and 3C281 do not show
a significant excess on the smaller scale, but do show a signal at
5-6$\sigma$ in the outer annulus, and of 2.5-3$\sigma$ overall. 

We use PIMMS to infer a luminosity for this extended emission from the
observed excess counts in the large annulus (1.5-50 arcsec), assuming that they are
due to simple thermal bremsstrahlung emission from gas at a
temperature of 4~keV at the same redshift as the quasar (except for
1821+643, where we use the measured value of 10~keV from Fang et al
2002), and that the light is subject only to Galactic absorption.
These luminosities, however, would exclude the very inner (brighter)
and far outer regions of any cluster. We correct for this factor by
assuming that the excess extended emission follows a beta model, with
$\beta=2/3$ and a core radius of 100\kpc, in order to estimate the
total luminosity of the cluster (ie out to a radius of 5 times the
core radius). For all the quasars the total luminosity of a cluster
following this profile is only a factor of $\sim1.2-1.3$ more than
that within the 1.5-50 arcsec annulus. We tabulate the total inferred
0.5-7~keV and bolometric luminosities of the surrounding cluster in
Table~\ref{tab:excess}.

\section{Discussion and Conclusions }

We clearly detect soft X-ray emission associated with a hotspot in one
of the lobes of the radio source in three of the 3C quasars. Where
seen in other strong radio sources at the centre of a cluster -- eg
Cygnus A (Wilson, Young \& Shopbell 2000), 3C295 (Harris et al 2000),
3C123 (Hardcastle, Birkinshaw \& Worrall 2001) -- such X-ray emission
is well-fit as synchrotron-self Compton emission from the
radio-emitting electrons with a magnetic field close to equipartition.
This interpretation is supported by the detailed analysis of the X-ray
emission from the hotspots and lobes of 3C263 by Hardcastle et al
(2002). The smoothed images shown in Figure~\ref{fig:ximgs} 
indicate that there is also jetted soft X-ray
emission associated with the radio source axis between the lobes,
similar to the inverse Compton
emission from lobes and jets in other intermediate-redshift quasars such as 3C207 
(Brunetti et al 2002) and 3C 351 (Hardcastle et al 2002). 

We find very significant evidence for soft, spatially extended
emission around the intermediate redshift quasars 3C263 and 3C275.1,
with inferred luminosities for this component of L$_{\rm Bol}$ =
$16.4$ and $7.6\times10^{43}$\ergps respectively. We also find
evidence for extended emission around
3C254 and 3C281, of L$_{\rm Bol}\sim3-4\times10^{43}$\ergps. We find
the excess of counts continues out to at least 50 arcsec,
corresponding to around 350--400~kpc at these redshifts. Both the
luminosities and the lengthscales of this extended X-ray emission
component match well the typical properties of the cores of clusters
of galaxies at low redshift.

Despite the conservative approach used in this paper by scaling and
subtracting the model PSF at 1.25~arcsec -- ie above a radius of
10~kpc for most of our quasars, whereas it is possible that the
cluster component may also contribute to the flux at the core of the
source -- our bolometric luminosity for the cluster around 1821+63
(assuming the observed temperature of 10~keV) agrees well with the
L$_{\rm Bol}\sim5\times10^{45}$\ergps given in Fang et al (2002).
Our value of $1.6\times10^{44}$\ergps for the bolometric
luminosity of the cluster around 3C263 is around half of the value of
$3\times10^{44}$\ergps  found by Hardcastle et al (2002). 

There is more discrepancy between our {\sl Chandra}-derived
luminosities for the extended component and those published
previously for the 3C quasars from {\sl ROSAT} observations (Crawford
et al 1999; Hardcastle \& Worrall 1999). First of all we note that
with {\sl Chandra} we are able to resolve out the X-ray emission
associated with the radio source and exclude it from the derived
surface brightness profile. An immediate comparison of the entries in
Tables~\ref{tab:hots} and \ref{tab:excess} shows that the unwitting
inclusion of the X-ray hotspots alone in the PSF would boost the
cluster luminosities we derive by 45, 25 and 45 per cent for for
3C254, 3C263 and 3C275.1 respectively. Thus correcting for the assumed
inclusion of the X-ray hotspot in the {\sl ROSAT} data, our value for
the extended component of 3C275.1 is similar to the {\sl ROSAT}
result (Crawford et al 1999). The cluster component around 3C263 was
not previously detected by either Hardcastle \& Worrall (1999) or Hall
et al (1995), although it is clearly present in the {\sl Chandra}
observation (see also Hardcastle et al 2002). The largest shortfalls in cluster luminosity 
compared to the {\sl ROSAT} values (Crawford et al 1999; Hardcastle
\& Worrall 1999) are for 3C254 and 3C281.
 We speculate that these differences only serve to highlight the
uncertainty in the {\sl ROSAT} results which could have been affected
by an insufficient or erroneous wobble correction, whereas the {\sl
Chandra} data are of superior resolution and quality. We note that a
decrease in the derived cluster luminosities from the {\sl ROSAT}
values would be more consistent with the estimates of gas pressure
derived from the ionization state of the extended emission-line
nebulae (Crawford \& Vanderriest 1997, 2000; Bremer et al 1992).


\section*{Acknowledgements}
We thank Kazushi Iwasawa for discussion about the {\sl Chandra} PSFs
at an early stage in this work, the CXC for provision of software and
the PSF library, and the referee for helpful comments. ACF and CSC
both thank the Royal Society for financial support.

{}

\onecolumn
\begin{table}
\caption{Log of the observations \label{tab:obs}}
\begin{tabular}{llccccc}
          & Name & Redshift &Exposure &N$_{\rm H}$ & Cts/frame time &Off-axis angle\\
          &      & $z$  &(ksec)   & ($10^{20}$\pcmsq) &  (\ctps)        & (arcmin) \\
1111+408 &  3C254     & 0.734 &29.67    &1.9       &0.615         &1.35\\
1137+660 &  3C263     & 0.646  &33.03  & 1.2          &0.646        &0.85\\
1241+166 & 3C275.1   & 0.555  & 24.76 & 2.0          & 0.284       & 0.64\\
1305+069 &  3C281     & 0.602  &15.85  & 2.2          &0.474        &1.29\\
1821+643 &           & 0.297  &99.62    & 4.0           &0.611         &0.14\\
\end{tabular} 
~\\
The Galactic absorption values given in the fifth column (N$_{\rm H}$) are estimated using the
hydrogen column densities from Stark et al (1992). \\
In the case of 3C263, only the usable time from the total exposure is
given in column 4.  \\
\end{table}

\begin{table}
\caption{Detections of X-ray emission from the radio hotspots \label{tab:hots}}
\begin{tabular}{lcccccccc}
Hotspot     & RA         & Dec         & Radius& 0.5-7 keV & 0.5-2 keV & 2-7 keV & L(0.5-2~keV) & L$_{\rm Bol}$ \\
            & (J2000)   &  (J2000)       & (arcsec) & (counts) & (counts) & (counts) & ($10^{42}$\ergps) & ($10^{42}$\ergps) \\ 
3C254 (W)    &11:14:37.75  &+40:37:22.88 &2 &21.3$\pm$4.7 &18.7$\pm$4.4 &2.6$\pm$1.7   & 5.7$\pm$1.3 & 35.2$\pm$8.3 \\
3C263 (SE)    &11:39:59.48  &+65:47:43.66 &2.5 &82.8$\pm$9.2 &68.4$\pm$8.3 &14.4$\pm$3.9  & 14.1$\pm$1.7 & 87.5$\pm$10.6 \\
3C275.1 (NW) &12:43:57.40 &+16:23:01.48 &2   &69.5$\pm$8.4 &55.8$\pm$7.5 &13.7$\pm$3.7  &  11.1$\pm$1.5 & 68.8$\pm$9.2 \\
3C281 (N)    &13:07:54.13 &+06:42:28.73 &5   & 4.0$^{\ast}$ & 4.0$^{\ast}$ & -- & 1.5$^{\ast}$ & 9.2$^{\ast}$ \\
\end{tabular}
~\\ The counts tabulated were extracted from a circular aperture
centred on the given RA and Dec, using the radius listed in the fourth column. \\ 
The luminosities have been estimated
from the soft (0.5-2~keV) countrates assuming that the spectrum of the
hotspot emission follows a $\Gamma=2$ power-law model, with only
Galactic absorption along the line of sight. \\
$^\ast$ A feature appears in the smoothed image
of 3C281 (Figure~\ref{fig:ximgs}) coincident with the radio lobe, but it is not
significant at 3~$\sigma$. \\
\end{table}


\begin{table}
\caption{Log of the observations used in empirical PSF determination \label{tab:emp}}
\begin{tabular}{lcrrrccl}
Name         &ObsID   &Exposure   &RA          &Dec      &Cts/frame-time & Off-axis angle & Source Description \\
             &        &(ksec)       &(J2000)     &(J2000)  & (\ctps) &(arcmin) & \\
CD-33~7795     &971     &9349  &11:31:55.24  &$-$34:36:27.11  &0.472 &0.33 & M3Ve T-Tau star \\
0235+164       &884 &27775 &02:38:38.96 &+16:36:59.94  &0.115  &0.62 & $z=0.94$ OVV QSO \\
PG1634+706$^{\ast}$ &-- &35202 &16:34:29.04 &+70:31:32.80 &0.189--0.266 &0.31-0.34 & $z=1.34$ QSO \\
PKS~2126-158    &376 &34132 &21:29:12.16 &--15:38:40.69  &0.645 &0.35 & $z=3.27$ QSO \\
\end{tabular}
~\\
Cts/frame-time are estimated from the total counts (in the full {\sl
Chandra} energy range) within a 10 arcsec radius aperture centred on
the source. \\
$^{\ast}$PG1634+706 is the merged image from 6 different observations,
accounting for the range in cts/frame-time and
off-axis angle given. \\

\end{table}

\begin{table}
\caption{Excess counts and inferred cluster luminosities. \label{tab:excess}}
\begin{tabular}{lrrrcccc}
               &   Excess counts [$\sigma$] &  Excess counts [$\sigma$] &  Excess counts [$\sigma$] & L(0.5-7 keV) & L$_{\rm Bol}$  \\
               &   (1.5-10$''$)             &  (10-50$''$)              & (1.5-50$''$)              &($10^{43}$\ergps) &($10^{43}$\ergps) \\
3C254          &  13.5$\pm$13.5  [12.9]     &  29.5$\pm$9.1 [6.0]       & 43.1$\pm$16.3  [14.2]     & 3.08$\pm$1.16 & 3.67$\pm$1.39 \\
3C263          &  137.5$\pm$23.4 [20.1]     &  129.1$\pm$14.6 [7.9]     & 266.6$\pm$27.6 [21.6]     & 13.34$\pm$1.38 & 16.36$\pm$1.69 \\
3C275.1        &  108.6$\pm$13.8 [8.3]      &  15.8$\pm$7.1   [3.1]     & 124.4$\pm$15.5 [8.9]      &  6.34$\pm$0.79 & 7.57$\pm$0.92 \\
3C281          &    7.1$\pm$12.0 [11.5]     &  25.5$\pm$8.6   [4.3]     &  32.6$\pm$14.7 [12.3]     &  3.03$\pm$1.37 & 3.59$\pm$1.89 \\
1821+643       & 1038.0$\pm$42.3 [27.2]     &  4142.4$\pm$72.0$^*$ [12.2] & 5180.4$\pm$83.5$^*$ [29.8] & 301.73$\pm$4.86$^\dag$ & 485.39$\pm$7.82$^\dag$ \\
\end{tabular}
~\\ The 0.5-2~keV counts in excess of the scaled model are given
summed over radii of 1.5-10 arcsec (second column), 10-50 arcsec
(third column)  and 1.5-50 arcsec
(fourth column). The exception (marked by an asterisk) is for 1821+643
where the excess counts in the third and fourth columns are from an
annulus reaching out to 75 arcsec. 
The 1-$\sigma$ uncertainties in the excess counts are also given. The
uncertainty on the predicted counts from the model PSF is listed in square brackets next to each value. \\
The luminosities given are the total cluster
luminosity, scaling up from the 0.5-2~keV counts from the 1.5-50 arcsec annulus
(fourth column); see text for details. The conversion assumes a 
thermal bremsstrahlung spectrum from gas at a
temperature of 4~keV at the redshift of the quasar.
The exception is again 1821+643 (luminosities marked by a $^\dag$)
where we have assumed a temperature of 10~keV as measured by Fang et
al (2002).  \\ \end{table}
\twocolumn

\onecolumn


\onecolumn

\twocolumn

\label{lastpage}


\begin{thebibliography}{}
\bibitem [] {} Bremer  MN, Crawford CS, Fabian AC, Johnstone RM, 1992, MNRAS, 254, 614
\bibitem [] {} Brunetti G, Bondi M, Comastri A, Setti G, 2002, A\&A, 381, 795 
\bibitem []{} Crawford CS, Lehmann I, Fabian AC, Bremer MN, Hasinger
G,  1999, MNRAS, 308, 1159
\bibitem []{} Crawford CS, Vanderriest C, 1997, MNRAS, 285, 580
\bibitem []{} Crawford CS, Vanderriest C, 2000, MNRAS, 315, 433
\bibitem [] {} Durret  F, Pecontal E, Petitjean P, Bergeron J, 1994, A\&A, 291, 392
\bibitem [] {} Ellingson E, Yee HKC, Green RF, 1991, ApJS, 76, 455
\bibitem []{} Fang T, Davis DS, Lee JC, Marshall HL, Bryan GL, Canizares CR, 2002, ApJ, 565, 86. 
\bibitem []{} Hall PB, Ellingson E, Green RF, Yee HKC, 1995, AJ, 110, 513
\bibitem []{} Hardcastle MJ, Worrall DM, 1999, MNRAS, 309, 969
\bibitem []{} Hardcastle MJ, Birkinshaw M, Worrall DM, 2001, MNRAS, 232, L17
\bibitem []{} Hardcastle MJ, Birkinshaw M, Cameron RA, Harris DE, Looney LW, Worrall DM, 2002 (astro-ph/0208204). 
\bibitem []{} Harris DE, Hjorth J, Sadun AC, Silverman JD, Vestergaard
M, 1999, ApJ, 518, 213
\bibitem []{} Hutchings JB, Dewey A, Chaytor D, Ryneveld S, Gower AC, Ellingson E, 1998, PASP, 100, 111
\bibitem []{} Owen FN, Puschell JJ, 1984, AJ, 89, 932 
\bibitem[]{} Sambruna R, Eracleous M, Mushotzky RF, 1999 ApJ, 526, 60
\bibitem []{} Stark AA, Gammie CF, Wilson RW, Bally J, Linke RA, Heiles C, Hurwitz M, 1992, ApJS, 79, 77
\bibitem []{} Wilson AS, Young AJ, Shopbell PL, 2000, ApJ 544, L27
\bibitem []{} Wold M, Lacy M, Lilje P, Serjeant S, 2000, MNRAS, 316, 267
\bibitem [] {} Yee HKC, Green RF, 1987, ApJ, 319, 28
              
\end{thebibliography}
\end{document}